\documentclass[12pt,a4paper]{article}
\usepackage{epsfig}
\usepackage{mathrsfs}

\def\hybrid{\topmargin -20pt    \oddsidemargin 0pt
        \headheight 0pt \headsep 0pt
        \textwidth 6.25in       
        \textheight 9.5in       
        \marginparwidth .875in
        \parskip 5pt plus 1pt   \jot = 1.5ex}

\hybrid

\def\baselinestretch{1.2}

\catcode`\@=11

\def\marginnote#1{}
\newcount\hour
\newcount\minute
\newtoks\amorpm
\hour=\time\divide\hour by60
\minute=\time{\multiply\hour by60 \global\advance\minute by-\hour}
\edef\standardtime{{\ifnum\hour<12 \global\amorpm={am}%
        \else\global\amorpm={pm}\advance\hour by-12 \fi
        \ifnum\hour=0 \hour=12 \fi
        \number\hour:\ifnum\minute<10 0\fi\number\minute\the\amorpm}}
\edef\militarytime{\number\hour:\ifnum\minute<10 0\fi\number\minute}

\def\draftlabel#1{{\@bsphack\if@filesw {\let\thepage\relax
   \xdef\@gtempa{\write\@auxout{\string
      \newlabel{#1}{{\@currentlabel}{\thepage}}}}}\@gtempa
   \if@nobreak \ifvmode\nobreak\fi\fi\fi\@esphack}
        \gdef\@eqnlabel{#1}}
\def\@eqnlabel{}
\def\@vacuum{}
\def\draftmarginnote#1{\marginpar{\raggedright\scriptsize\tt#1}}

\def\draft{\oddsidemargin -.5truein
        \def\@oddfoot{\sl preliminary draft \hfil
        \rm\thepage\hfil\sl\today\quad\militarytime}
        \let\@evenfoot\@oddfoot \overfullrule 3pt
        \let\label=\draftlabel
        \let\marginnote=\draftmarginnote
   \def\@eqnnum{(\theequation)\rlap{\kern\marginparsep\tt\@eqnlabel}%
\global\let\@eqnlabel\@vacuum}  }

\def\preprint{\twocolumn\sloppy\flushbottom\parindent 2em
        \leftmargini 2em\leftmarginv .5em\leftmarginvi .5em
        \oddsidemargin -.5in    \evensidemargin -.5in
        \columnsep .4in \footheight 0pt
        \textwidth 10.in        \topmargin  -.4in
        \headheight 12pt \topskip .4in
        \textheight 6.9in \footskip 0pt
        \def\@oddhead{\thepage\hfil\addtocounter{page}{1}\thepage}
        \let\@evenhead\@oddhead \def\@oddfoot{} \def\@evenfoot{} }

\def\numberbysection{\@addtoreset{equation}{section}
        \def\theequation{\thesection.\arabic{equation}}}

\def\underline#1{\relax\ifmmode\@@underline#1\else
        $\@@underline{\hbox{#1}}$\relax\fi}

\def\titlepage{\@restonecolfalse\if@twocolumn\@restonecoltrue\onecolumn
     \else \newpage \fi \thispagestyle{empty}\c@page\z@
        \def\thefootnote{\fnsymbol{footnote}} }

\def\endtitlepage{\if@restonecol\twocolumn \else \newpage \fi
        \def\thefootnote{\arabic{footnote}}
        \setcounter{footnote}{0}}  

\catcode`@=12
\relax

\def\figcap{\section*{Figure Captions\markboth
        {FIGURECAPTIONS}{FIGURECAPTIONS}}\list
        {Figure \arabic{enumi}:\hfill}{\settowidth\labelwidth{Figure
999:}
        \leftmargin\labelwidth
        \advance\leftmargin\labelsep\usecounter{enumi}}}
 \relax
\def\tablecap{\section*{Table Captions\markboth
        {TABLECAPTIONS}{TABLECAPTIONS}}\list
        {Table \arabic{enumi}:\hfill}{\settowidth\labelwidth{Table
999:}
        \leftmargin\labelwidth
        \advance\leftmargin\labelsep\usecounter{enumi}}}
 \relax
\def\reflist{\section*{References\markboth
        {REFLIST}{REFLIST}}\list
        {[\arabic{enumi}]\hfill}{\settowidth\labelwidth{[999]}
        \leftmargin\labelwidth
        \advance\leftmargin\labelsep\usecounter{enumi}}}
 \relax
\makeatletter
\newcounter{pubctr}
\def\publist{\@ifnextchar[{\@publist}{\@@publist}}
\def\@publist[#1]{\list
        {[\arabic{pubctr}]\hfill}{\settowidth\labelwidth{[999]}
        \leftmargin\labelwidth
        \advance\leftmargin\labelsep
        \@nmbrlisttrue\def\@listctr{pubctr}
        \setcounter{pubctr}{#1}\addtocounter{pubctr}{-1}}}
\def\@@publist{\list
        {[\arabic{pubctr}]\hfill}{\settowidth\labelwidth{[999]}
        \leftmargin\labelwidth
        \advance\leftmargin\labelsep
        \@nmbrlisttrue\def\@listctr{pubctr}}}
 \relax
\makeatother
\newskip\humongous \humongous=0pt plus 1000pt minus 1000pt

\newif\ifdtup

\relax

\def\be{\begin{equation}}
\def\ee{\end{equation}}
\def\ba{\begin{eqnarray}}
\def\ea{\end{eqnarray}}

\def\no{\noindent}

\def\IR{\relax{\rm I\kern-.18em R}}

\begin{document}

\renewcommand{\theequation}{\thesection.\arabic{equation}}

\newcommand{\beq}{\begin{equation}}
\newcommand{\eeq}[1]{\label{#1}\end{equation}}
\newcommand{\ber}{\begin{eqnarray}}
\newcommand{\eer}[1]{\label{#1}\end{eqnarray}}
\newcommand{\eqn}[1]{(\ref{#1})}
\begin{titlepage}
\begin{center}

\hfill October 2009\\

\vskip .4in

{\large \bf Dual photons and gravitons}\footnote{Based on
lectures delivered at {\em ``String Theory and Fundamental
Physics"}, Kanha, India, 11-17 February 2009; {\em ``Spring
School in Strings, Cosmology and Particles"}, Ni\v{s}, Serbia,
31 March - 4 April 2009 ; {\em ``IPM String School and Workshop"},
Tehran, Iran, 9-18 April 2009; XXXIX\`eme Institut d'\'et\'e
{\em ``AdS, TCC et Probl\`emes Apparent\'es"}, Paris, France,
17-28 August 2009. Contribution to appear in
a special issue in Monographs Series of the Publications of
the Astronomical Observatory of Belgrade.}

\vskip 0.6in

{\bf Ioannis Bakas}
\vskip 0.2in
{\em Department of Physics, University of Patras \\
GR-26500 Patras, Greece\\
\footnotesize{\tt bakas@ajax.physics.upatras.gr}}\\

\end{center}

\vskip .8in

\centerline{\bf Abstract}

\no
We review the status of electric/magnetic duality for free gauge field
theories in four space-time dimensions with emphasis on Maxwell theory
and linearized Einstein gravity. Using the theory of vector and tensor
spherical harmonics, we provide explicit construction of dual photons
and gravitons by decomposing the fields into axial and polar configurations
with opposite parity and interchanging the two sectors. When the theories
are defined on $AdS_4$ space-time there are boundary manifestations
of the duality, which for the case of gravity account for the
energy-momentum/Cotton tensor duality (also known as dual graviton
correspondence). For $AdS_4$ black-hole backgrounds there is no direct
analogue of gravitational duality on the bulk, but there is still a
boundary duality for quasi-normal modes satisfying a selected set of
boundary conditions. Possible extensions of this framework and
some open questions are also briefly discussed.
\vfill
\end{titlepage}
\eject

\def\baselinestretch{1.2}
\baselineskip 16 pt
\noindent


\section{Introduction}
\setcounter{equation}{0}

Electric/magnetic duality has been the cornerstone for many developments
in theoretical and mathematical physics in the past fifteen years.
It has its roots in Dirac's original work on electromagnetism, but it
certainly goes well beyond it in many ways. In recent years it has also been
generalized to other free gauge field theories including linearized gravity
in four space-time dimensions. Some implications of electric/magnetic duality
have also been studied in the context of $AdS_4/CFT_3$ correspondence, while
several other ideas related to duality in gravitational theories are still
waiting to find their proper place in this framework and (hopefully) turn into
powerful tools.

The purpose of these lecture notes is to review the basic elements of duality
in electromagnetism and gravity and resolve the non-local relations that
define the dual field variables in terms of the original ones. This is practically
achieved by employing group theory techniques for solving wave equations on
spherically symmetric backgrounds. The main emphasis here is on the methods 
that lead to the resolution of dualities, whereas the actual details of the
calculations are omitted in some cases to avoid lengthy formulae; the interested
reader will be referred to the literature for further technicalities.
We found it rather instructive to present these methods in parallel for Maxwell
theory and linearized gravity because of the common aspects they share. Our
treatment of the problem for Maxwell equations is quite elementary, but is not
usually found elsewhere. Thus, we include it here for completeness and in
order to motivate in pedagogical way similar constructions for linearized gravity.
As application of all these results, we discuss the boundary manifestation
of electric/magnetic duality for theories defined on $AdS_4$ space-time and
explain, in particular, how the dual graviton correspondence comes into
play in holography. Lack of space does not allow us to expand further on these
applications; here, we only give a flavor of the additional features one
has to account and understand better in the context of $AdS_4/CFT_3$
correspondence.

Thus, the material of this contribution is organized as follows. Section 2
provides a brief outline of the electric/magnetic duality in Maxwell theory and
linearized gravity together with the definition of dual photons and gravitons.
Section 3 contains some background mathematical material from the theory
of spherical harmonics, which are used to decompose gauge fields as well as
metric perturbations into the axial and polar sectors with opposite parity.
Section 4 makes use of the classical equations of motion, thus reducing the
corresponding wave equations into an effective Schr\"odinger problem.
Then, duality is resolved group theoretically by exchanging the axial and
polar sectors of the theories and leads to explicit construction of the
dual photons and gravitons. Section 5 makes use of holographic renormalization
to examine the boundary manifestation of electric/magnetic duality on
$AdS_4$ space-time for both Maxwell theory and linearized gravity. Certain
generalizations to $AdS_4$ black-hole backgrounds are also briefly discussed
in this context. Finally, section 6 contains our conclusions and discusses
possible extensions of the formalism to address some open problems.

Further details can be found in the published works \cite{bakas1}, \cite{bakas},
and references therein.

\section{Electric/magnetic duality}
\setcounter{equation}{0}

Free gauge fields in four space-time dimensions exhibit two physical degrees
of freedom that can be rotated into one another by a canonical transformation
mixing the two pairs of unconstrained dynamical variables, while
keeping the Hamiltonian form-invariant. It is a general result that extends the
electric/magnetic duality of electromagnetism to other physical fields
including Einstein gravity. In this section we briefly review the status of
duality in Maxwell and Einstein theories and provide the definition of
dual photons and dual gravitons. Their explicit construction will be described
in detail in subsequent sections.

\subsection{Maxwell theory}

Let us consider the source free Maxwell theory on a fixed (generally curved)
space-time manifold $M_4$ described in terms of a $U(1)$ gauge field $A_{\mu}$.
The field strength is defined as usual,
\be
F_{\mu \nu} = \nabla_{\mu} A_{\nu} - \nabla_{\nu} A_{\mu} ~,
\ee
and the dual field strength as
\be
{}^{\star}F_{\mu \nu} = {1 \over 2} {\epsilon_{\mu \nu}}^{\kappa \lambda}
F_{\kappa \lambda}
\ee
using the covariant fully antisymmetric symbol on $M_4$.

The field equations and the Bianchi identities assume the familiar form
\be
\nabla_{\nu} F^{\mu \nu} = 0 = \nabla_{\nu} {}^{\star}F^{\mu \nu}
\ee
and their role is interchanged under the operation of duality
satisfying the identity $\star^2 = -1$. When expressed in terms of physical
fields,
\be
E_a = F_{ta} ~, ~~~~~ B_a = {}^{\star}F_{ta}
\ee
it interchanges electric and magnetic components as follows,
\be
E_a \rightarrow B_a ~, ~~~~~~ B_a \rightarrow - E_a ~,
\ee
thus giving rise to the celebrated electric/magnetic duality of Maxwell theory.

Then, for a given gauge field $A_{\mu}$, the dual configuration $\tilde{A}_{\mu}$
is defined by the equations
\ba
& & {}^{\star}F_{\mu \nu} (A) = F_{\mu \nu} (\tilde{A}) ~, \\
& & {}^{\star}F_{\mu \nu} (\tilde{A}) = - F_{\mu \nu} (A) ~.
\ea
$\tilde{A}_{\mu}$, which we call it the dual photon, is uniquely defined, up
to gauge transformations, but it is non-locally related to the original field
$A_{\mu}$. It is our purpose to provide explicit construction of the dual
field in a broad class of static spherically symmetric backgrounds $M_4$.
It will also prove a useful guide for performing similar constructions in
linearized gravity around (a more restricted class of) spherically
symmetric backgrounds.

We also note for completeness that one may consider an $SO(2)$ rotation of the
electric and magnetic fields parametrized by an arbitrary angle $\delta$,
\be
\left(\begin{array}{c}
E_a^{\prime} \\
                       \\
B_a^{\prime}
\end{array} \right) =
\left(\begin{array}{ccc}
{\rm cos} \delta &  & {\rm sin} \delta \\
                 &  &                \\
- {\rm sin} \delta  &  & {\rm cos} \delta
\end{array} \right)
\left(\begin{array}{c}
E_a \\
                       \\
B_a
\end{array} \right) ,
\ee
thus providing a more general action on the space of solutions of Maxwell
theory.

\subsection{Linearized gravity}

For gravity, the electric/magnetic duality is only defined at the linear
level by considering small perturbations around a reference metric,
\be
g_{\mu \nu} = g_{\mu \nu}^{(0)} + h_{\mu \nu} ~.
\ee
Then, duality is described  as a non-local transformation among perturbations
of the same reference metric
\be
\tilde{g}_{\mu \nu} = g_{\mu \nu}^{(0)} + \tilde{h}_{\mu \nu} ~,
\ee
which acts on the space of solutions of vacuum Einstein equation and which
can also be realized as symmetry of the gravitational
action; the duality breaks down at the first self-interacting
cubic approximation to general relativity, but this will not be relevant
to the present work. The actual differential equations that define the
dual graviton $\tilde{h}_{\mu \nu}$ in terms of $h_{\mu \nu}$ resemble those of
Maxwell theory and will be written down shortly.

We mention right from the beginning that the reference metric $g_{\mu \nu}^{(0)}$
cannot be arbitrary, and, hence, gravitational duality (as known to this day) is only
restricted to small perturbations around special backgrounds.
These are provided by the metric of flat Minkowski space-time
when the cosmological constant $\Lambda = 0$ and by the metric of
$(A)dS_4$ space-time when $\Lambda \neq 0$. The reason for it, as will emerge from
the formalism, is that these special backgrounds are trivially self-dual (in an
appropriate sense) and they remain inert under duality. Then, one is only left to
compare the fluctuations $h_{\mu \nu}$ and establish the action of the duality
symmetry.

The electric/magnetic duality of linearized gravity was initially formulated
for $\Lambda = 0$, but it extends rather easily to vacuum Einstein equations with
cosmological constant, \cite{nieto}, \cite{hull}, \cite{claudio}, \cite{stan1},
\cite{julia}, \cite{tassos}, \cite{bakas1}. Here, following ref. \cite{bakas1},
we formulate directly the problem for arbitrary $\Lambda$ using the quantity
\be
Z_{\mu \nu \rho \sigma} = R_{\mu \nu \rho \sigma} - {\Lambda \over 3}
\left(g_{\mu \rho} g_{\nu \sigma} - g_{\mu \sigma} g_{\nu \rho} \right) ~.
\ee
$Z_{\mu \nu \rho \sigma}$ arises by restricting the Weyl curvature tensor in four
space-time dimensions to on-shell metrics, and, clearly, it fulfills the identities
\be
Z_{\mu [\nu \rho \sigma ]} = 0 ~, ~~~~~
\nabla_{[ \lambda} Z_{\mu \nu ] \rho \sigma} = 0 ~,
\ee
and the on-shell metrics satisfy the equation
\be
{Z^{\rho}}_{\mu \rho \nu} \equiv Z_{\mu \nu} = 0 ~,
\ee
which is equivalent to Einstein equations with cosmological constant
$\Lambda$, i.e., $R_{\mu \nu} = \Lambda g_{\mu \nu}$.
One also defines the dual curvature tensor
\be
{}^{\star}Z_{\mu \nu \rho \sigma} = {1 \over 2}
{\epsilon_{\mu \nu}}^{\kappa \lambda} Z_{\kappa \lambda \rho \sigma} ~,
\ee
which fulfills similar identities, but with reverse meaning.

Linearized gravity around $g_{\mu \nu}^{(0)}$ exhibits a duality that
exchanges Bianchi identities with the classical equations of motion,
as in electromagnetism, by letting
\ba
& & {}^{\star}Z_{\mu \nu \rho \sigma}(g) = Z_{\mu \nu \rho \sigma}
(\tilde{g}) ~, \\
& & {}^{\star}Z_{\mu \nu \rho \sigma}(\tilde{g}) = -
Z_{\mu \nu \rho \sigma} (g) ~.
\ea
Actually, at the linear level, it is appropriate to replace the covariant
derivatives $\nabla$ by ordinary derivatives $\partial$ and work out the
linear differential equations that uniquely define $\tilde{h}_{\mu \nu}$
in terms of $h_{\mu \nu}$, up to reparametrizations. This is in exact
analogy with electromagnetism.

Also, it is useful to introduce the electric and magnetic components
of the Weyl tensor as
\be
E_{ab} = Z_{a t b t} ~, ~~~~~ B_{ab} =
{}^{\star}Z_{a t b t} ~.
\ee
Then, the gravitational duality transformation is realized as
\be
E_{ab} \rightarrow B_{ab} ~, ~~~~~
B_{ab} \rightarrow - E_{ab} ~.
\label{dual}
\ee
The electric and magnetic tensors are represented on shell by $3 \times 3$
symmetric traceless matrices, and, as such, they have five
independent components each. Furthermore, the background metric $g_{\mu \nu}^{(0)}$,
which is used to exhibit duality for the appropriate choice of $\Lambda$, is
trivially self-dual, since it has $E_{ab} = 0 = B_{ab}$.

More generally, as for electromagnetism, one may consider an $SO(2)$ rotation of these
components parametrized by an arbitrary angle $\delta$,
\be
\left(\begin{array}{c}
E_{ab}^{\prime} \\
                       \\
B_{ab}^{\prime}
\end{array} \right) =
\left(\begin{array}{ccc}
{\rm cos} \delta &  & {\rm sin} \delta \\
                 &  &                \\
- {\rm sin} \delta  &  & {\rm cos} \delta
\end{array} \right)
\left(\begin{array}{c}
E_{ab} \\
                       \\
B_{ab}
\end{array} \right) ,
\ee
thus providing a more general action on the space of solutions of linearized
gravity. Details of the proof can be found in the original works (see, in
particular, ref. \cite{claudio} and \cite{julia}).

Our purpose in the sequel is to provide explicit construction of the dual
gravitons, in analogy with the dual photons, by relying on group theory
techniques for solving wave equations on spherically symmetric space-times.

\section{Analysis into spherical harmonics}
\setcounter{equation}{0}

We consider static spherically symmetric space-times $M_4$ with local coordinates
$(t, ~ r, ~ \theta , ~ \phi)$ and line element
\be
ds^2 = -f(r) dt^2 + {dr^2 \over f(r)} + r^2 (d \theta^2 + {\rm sin}^2 \theta
d \phi^2) ~.
\label{themetric}
\ee
Wave equations for physical fields, such as a gauge field $A_{\mu}$ satisfying
the source-free Maxwell equations on $M_4$, and a graviton field $h_{\mu \nu}$
satisfying the linearized Einstein equations for metric perturbations around $M_4$,
can be reduced to a radial (effective) Schr\"odinger equation by assuming
factorization of their solutions on the $(t, ~ r)$ and $(\theta , ~ \phi)$
coordinates. In this section, we present the background mathematical material to
achieve this purpose for fields defined on $M_4$.

\subsection{Generalities}

The dependence of fields upon $\theta$ and $\phi$ is solely described
by group theory using spherical harmonics and generalizations thereof. In particular,
one should consider the theory of spherical harmonics for vector and
tensor fields, generalizing the usual spherical harmonics $Y_l^m (\theta, \phi)$
for scalar fields, to higher rank configurations.

The spherical symmetry of $M_4$ allows angular momentum to be defined and investigated
by studying rotations on the two-dimensional manifold having $t = {\rm constant}$ and
$r = {\rm constant.}$ Thus, to set up the notation, we
consider the sphere $S^2$ with local coordinates $(\theta, ~ \phi)$ and metric tensor
\be
\gamma_{ij} = \left(\begin{array}{ccc}
1 &  & 0 \\
  &  &   \\
0 &  & {\rm sin}^2 \theta
\end{array} \right) .
\ee
We also define the covariant fully antisymmetric symbol on $S^2$ with non-vanishing
components
\be
\epsilon^{\theta \phi} = - \epsilon^{\phi \theta} = {1 \over {\rm sin} \theta} ~.
\ee

Under rigid rotations around the origin, the various components of physical fields
defined on $M_4$ are decomposed into spherical harmonics by viewing them as covariant
quantities on the sphere $S^2$. Then, the theory of generalized spherical harmonics
(see, for instance, ref. \cite{wheeler}, \cite{zeril} and the review article \cite{thorne}) can be
tabulated as follows, for scalar, vector and rank-two tensor fields defined on $S^2$,
all having definite parity:

\underline{Scalar}: They are simply described by ordinary spherical harmonics with
$l = 0, 1, 2, \cdots$ and $-l \leq m \leq l$,
\be
Y_l^m ~, ~~~~~~~ {\rm with ~ parity} ~ (-1)^l ~.
\ee

\underline{Vectors}: There are two different type of vector spherical harmonics with 
opposite parity, namely
\ba
(\Psi_l^m)_i & = & \partial_i Y_l^m ~, ~~~~~~~~~~ {\rm with ~ parity} ~ (-1)^l ~, \\
(\Phi_l^m)_{i} & = & {\epsilon_i}^j \partial_j Y_l^m ~,
~~~~~~~ {\rm with ~ parity} ~ (-1)^{l+1} ~.
\ea

\underline{Rank-two tensors}: In this case there are three different type of tensor
spherical harmonics with the following parities,
\ba
(\Psi_l^m)_{ij} & = & \nabla_i \nabla_j Y_l^m ~, ~~~~~~~~
{\rm with ~ parity} ~ (-1)^l ~, \\
(\Phi_l^m)_{ij} & = & \gamma_{ij} Y_l^m ~,
~~~~~~~~~~~~ {\rm with ~ parity} ~ (-1)^{l} ~, \\
(X_l^m)_{ij} & = & {1 \over 2} \Big[{\epsilon_i}^k (\Psi_l^m)_{kj}
+ {\epsilon_j}^k (\Psi_l^m)_{ki} \Big] ~, ~~~
{\rm with ~ parity} ~ (-1)^{l+1} ~.
\ea

In the examples of gauge fields $A_{\mu}$ and rank-two tensor fields
$h_{\mu \nu}$ on $M_4$ that will be considered below in detail, all scalar, vector and
tensor harmonics come naturally into play. In all cases, it is particularly convenient to
separate the fields into two distinct classes with definite, but opposite parity, since
the general configuration is a linear superposition of them. The {\em polar} class 
corresponds to fields of parity $(-1)^l$, whereas the {\em axial} class to parity 
$(-1)^{l+1}$. They can be though as the ``electric" and ``magnetic" parts of the fields,
respectively, due to their parity, and, as such, they provide the right decomposition
to realize electric-magnetic duality in a natural way. Clearly, this decomposition is
invariant under rigid rotations, since the parity does not change.

We also note that the radial equations that result from wave equations in
spherically symmetric space-times depend only on the angular momentum quantum number
$l$ and not on $m$, as in the hydrogen atom. Thus, without loss of generality, we may
drop the $\phi$ dependence by specializing to axisymmetric field configurations with $m=0$,
so that $Y_l^m (\theta , \phi)$ reduce to $P_l ({\rm cos} \theta)$, to simplify the
derivation of the radial equations described in the subsequent sections. This will 
be discussed further shortly, following the general harmonic decomposition of the fields.

One may similarly consider the harmonic decomposition of higher rank tensor fields,
but such generalization will not be in focus in the present work, since we only consider 
the theories of photons and gravitons.

\subsection{Gauge field decomposition}

Under rigid rotations around the origin, the four components of a gauge field $A_{\mu}$
on $M_4$ transform like two scalars $A_t$, $A_r$ and one vector $(A_{\theta}$, $A_{\phi}$), 
when considered as covariant quantities on the sphere $S^2$. Thus, it
is natural to use a $2+2$ splitting of the space-time components of $A_{\mu}$, as
\be
A_{\mu} (t, r, \theta , \phi) = \left(\begin{array}{c}
S \\
  \\
V
\end{array} \right) ,
\ee
denoting by (S) the scalars and by (V) the vector. Then, any $A_{\mu}$ can be decomposed
into two distinct classes of opposite parity with respect to angular momentum (see, for
instance, \cite{ruffini}):

\underline{Axial gauge field}: For gauge field configurations of parity $(-1)^{l+1}$ the
two scalars obviously vanish,
\be
S_{\rm axial} = \left(\begin{array}{c}
0 \\
  \\
0
\end{array} \right) ,
\ee
and the vector part is dictated by the form of the corresponding vector spherical harmonic
$(\Phi_l^m)_i$, using an arbitrary multiplicative function $a(t, r)$,  as
\be
V_{\rm axial} = a(t, r) \left(\begin{array}{c}
- {1 \over {\rm sin} \theta} {\partial \over \partial \phi} \\
  \\
{\rm sin} \theta {\partial \over \partial \theta}
\end{array} \right) Y_l^m (\theta, \phi) ~.
\ee

For axisymmetric configurations, with $m=0$, the most general gauge field with
parity $(-1)^{l+1}$ takes the form
\be
A_{\mu}^{\rm axial} = \left(\begin{array}{c}
0 \\
0\\
0\\
a(r)
\end{array} \right) e^{-i \omega t} {\rm sin} \theta ~ \partial_{\theta}
P_l({\rm cos} \theta)
\label{axvec}
\ee
by also factorizing the $t$ and $r$ dependence of the function $a(t, r)$.

\underline{Polar gauge field}: For gauge field configurations of parity $(-1)^l$
the scalar part is parametrized in terms of two arbitrary functions $C(t, r)$ and
$D(t, r)$ as
\be
S_{\rm polar} = \left(\begin{array}{c}
C (t, r) \\
  \\
D (t, r)
\end{array} \right) Y_l^m (\theta, \phi) ,
\ee
whereas the vector part is expressed in terms of the corresponding vector
spherical harmonic $(\Psi_l^m)_i$ as
\be
V_{\rm polar} = b(t, r) \left(\begin{array}{c}
{\partial \over \partial \theta} \\
  \\
{\partial \over \partial \phi}
\end{array} \right) Y_l^m (\theta, \phi) ~,
\ee
using an arbitrary function $b(t, r)$. Note, however, that the
vector part can always be gauged away to zero by the transformation
\be
A_{\mu}^{\prime} = A_{\mu} + \partial_{\mu} \epsilon
\ee
choosing, in particular, the parameter
\be
\epsilon = - b(r, t) Y_l^m (\theta, \phi) ~.
\ee
Thus, only the scalar part is non-vanishing in this case.

For axisymmetric configurations, with $m=0$, the most general gauge field with
parity $(-1)^l$ takes the form
\be
A_{\mu}^{\rm polar} = \left(\begin{array}{c}
C(r) \\
D(r)\\
0\\
0
\end{array} \right) e^{-i \omega t} P_l({\rm cos} \theta)
\label{povec}
\ee
by also factorizing the $t$ and $r$ dependence of the functions $C(t, r)$ and 
$D(t, r)$.

In the sequel, we assume without lose of generality that the gauge fields
$A_{\mu}$ are decomposed invariantly into axial and polar parts given by the
axisymmetric expressions \eqn{axvec} and \eqn{povec}. The equations for the
corresponding radial functions $a(r)$ and $C(r)$, $D(r)$ follow from the
Maxwell equations in vacuum, as will be seen later.

\subsection{Metric decomposition}

Under rigid rotations around the origin, the ten components of the tensor field $h_{\mu \nu}$
on $M_4$ transform like three scalars $h_{tt}$, $h_{tr}$, $h_{rr}$, two vectors
$(h_{t \theta}$, $h_{t, \phi})$ and $(h_{r \theta}$, $h_{r \phi})$, and a rank-two tensor
(accounting for the remaining components),
when considered as covariant quantities on the sphere $S^2$. Thus, it
is natural to use a $2+2$ splitting of the symmetric matrix $h_{\mu \nu}$ into blocks
\be
h_{\mu \nu} (t, r, \theta , \phi) = \left(\begin{array}{ccc}
S & & V \\
  & &   \\
\tilde{V} & & T
\end{array} \right)
\ee
denoting by (S) the scalars by (V) the vectors and by (T) the tensor, whereas $\tilde{V}$ denotes
the transpose of the $2 \times 2$ matrix $V$. Then, any $h_{\mu \nu}$ can
be decomposed into two distinct classes of opposite parity with respect to angular momentum,
\cite{wheeler}:

\underline{Axial metric perturbations}: For metric perturbations of parity $(-1)^{l+1}$ the scalar
part obviously vanishes,
\be
S_{\rm axial} = \left(\begin{array}{ccc}
0 &  & 0 \\
  &  &   \\
0 &  & 0
\end{array} \right) .
\ee
The vector part contains two vector spherical harmonics $(\Phi_l^m)_i$ with parity $(-1)^{l+1}$,
each one multiplied with an arbitrary function $h_0 (t, r)$ and $h_1 (t, r)$, and
takes the form
\be
V_{\rm axial} = \left(\begin{array}{cccc}
-h_0 (t, r) {1 \over {\rm sin} \theta} {\partial \over \partial \phi} &  & &
h_0 (t, r) {\rm sin} \theta {\partial \over \partial \theta} \\
   &   &  &  \\
   &   &  &  \\
-h_1 (t, r) {1 \over {\rm sin} \theta} {\partial \over \partial \phi} &  & &
h_1 (t, r) {\rm sin} \theta {\partial \over \partial \theta}
\end{array} \right) Y_l^m (\theta, \phi) ~.
\ee
Finally, the tensor part is obtained from the corresponding rank-two tensor spherical
harmonic $(X_l^m)_{ij}$ by multiplying it with an arbitrary function $h_2 (t, r)$,
\be
T_{\rm axial} = {1 \over 2} h_2 (t, r) \left(\begin{array}{cccc}
2 \left({1 \over {\rm sin} \theta} {\partial^2 \over \partial \theta
\partial \phi} - {{\rm cos} \theta \over {\rm sin}^2 \theta} {\partial \over \partial
\phi} \right) &  &  & {1 \over {\rm sin} \theta}
{\partial^2 \over \partial \phi^2} - {\rm sin} \theta {\partial^2 \over \partial
\theta^2} + {\rm cos} \theta {\partial \over \partial \theta} \\
  &  &  &  \\
  &  &  &  \\
{1 \over {\rm sin} \theta}
{\partial^2 \over \partial \phi^2} - {\rm sin} \theta {\partial^2 \over \partial
\theta^2} + {\rm cos} \theta {\partial \over \partial \theta} & & &
-2 \left({\rm sin} \theta {\partial^2 \over \partial \theta
\partial \phi} - {\rm cos} \theta {\partial \over \partial \phi} \right)
\end{array} \right) Y_l^m (\theta, \phi) ~.
\ee

Note, however, that one may employ reparametrizations generated by a vector field
$\xi_{\mu}$,
\be
h_{\mu \nu}^{\prime} = h_{\mu \nu} + \nabla_{\mu} \xi_{\nu} + \nabla_{\nu} \xi_{\mu} ~,
\ee
in order to simplify the expressions. For example, the terms containing second order
derivatives in $\theta$ and $\phi$ can be gauged away using
\be
\xi_{\mu} = \Lambda (t, r) \left(0, ~ 0, ~ - {1 \over {\rm sin} \theta}
{\partial \over \partial \phi} , ~ {\rm sin} \theta {\partial \over \partial \theta} \right)
Y_l^m (\theta, \phi)
\ee
with appropriately chosen function $\Lambda (t, r)$ to also annul $h_2 (t, r)$; this 
$\xi_{\mu}$ has exactly the same form as an axial gauge field with parity $(-1)^{l+1}$.

For axisymmetric configurations, with $m=0$, the general form of metric perturbations with
parity $(-1)^{l+1}$ becomes, after taking into account the freedom of reparametrizations,
\be
h_{\mu \nu}^{\rm axial} = \left(\begin{array}{cccc}
0 & 0 & 0 & h_0(r) \\
  &   &   &   \\
0 & 0 & 0 & h_1(r) \\
  &   &   &   \\
0 & 0 & 0 & 0 \\
  &   &   &   \\
h_0(r) & h_1(r) & 0 & 0
\end{array} \right)
e^{-i\omega t} {\rm sin} \theta ~
\partial_{\theta} P_l ({\rm cos} \theta)
\label{axmet}
\ee
by also factorizing the $t$ and $r$ dependence of the functions $h_0 (t, r)$ and
$h_1 (t, r)$.

\underline{Polar metric perturbations}: For metric perturbations of parity $(-1)^l$ the
scalar part is parametrized by 3 arbitrary functions $H_0 (t, r)$, $H_1 (t, r)$ and
$H_2 (t, r)$, one for each scalar,
\be
S_{\rm polar} = \left(\begin{array}{ccc}
f(r) H_0 (t, r) &  & H_1 (t, r) \\
  &  &   \\
  &  &   \\
H_1 (t, r) &  & H_2 (t,r) /f(r)
\end{array} \right) Y_l^m (\theta, \phi) ~.
\ee
The vector part is described by two vector spherical harmonics $(\Psi_l^m)_i$ of the
corresponding parity, using two arbitrary multiplicative functions $h_0 (t, r)$ and
$h_1 (t, r)$,
\be
V_{\rm polar} = \left(\begin{array}{cccc}
h_0 (t, r) {\partial \over \partial \theta} &  & &
h_0 (t, r) {\partial \over \partial \phi} \\
   &   &  &  \\
   &   &  &  \\
h_1 (t, r) {\partial \over \partial \theta} &  & &
h_1 (t, r) {\partial \over \partial \phi}
\end{array} \right) Y_l^m (\theta, \phi) ~.
\ee
Finally, the tensor part is composed out of two rank-two tensor spherical harmonics
$(\Psi_l^m)_{ij}$ and $(\Phi_l^m)_{ij}$ of parity $(-1)^l$ with arbitrary coefficients
$G(t, r)$ and $K(t, r)$, respectively, and it takes the
following form,
\ba
T_{\rm polar} & = & r^2 G(t, r) \left(\begin{array}{cccc}
{\partial^2 \over \partial \theta^2} &  &  &
{\partial^2 \over \partial \theta \partial \phi} -
{\rm cot} \theta {\partial \over \partial \phi}  \\
   &  &  &  \\
   &  &  &  \\
{\partial^2 \over \partial \theta \partial \phi} -
{\rm cot} \theta {\partial \over \partial \phi}  &  &  &
{\partial^2 \over \partial \phi^2} + {\rm sin} \theta {\rm cos} \theta
{\partial \over \partial \theta}
\end{array} \right) Y_l^m (\theta, \phi) \nonumber \\ \nonumber \\
& & + r^2 K(t, r) \left(\begin{array}{ccc}
1 & & 0 \\
  & &   \\
0 & & {\rm sin}^2 \theta
\end{array} \right) Y_l^m (\theta, \phi) ~.
\ea

As before, there is also the freedom of making reparametrizations generated by a vector
field $\xi_{\mu}$,
\be
h_{\mu \nu}^{\prime} = h_{\mu \nu} + \nabla_{\mu} \xi_{\nu} + \nabla_{\nu} \xi_{\mu} ~,
\ee
which can be chosen to gauge away the terms containing second derivatives in $\theta$ and
$\phi$. A particularly useful choice that serves this purpose is provided by
\be
\xi_{\mu} = \left(M_0 (t, r) , ~ M_1 (t, r) , ~ M(t, r) {\partial \over \partial \theta}
, ~ M(t, r) {\partial \over \partial \phi} \right)
Y_l^m (\theta, \phi)
\ee
with appropriately chosen functions $M_0 (t, r)$, $M_1 (t, r)$ and $M(t, r)$ to also annul 
$h_0 (t, r)$, $h_1 (t, r)$ and $G(t, r)$; here, $\xi_{\mu}$
has exactly the same form as a polar gauge field with parity $(-1)^l$.

For axisymmetric configurations, with $m=0$, the general form of metric perturbations with
parity $(-1)^l$ becomes, after taking into account the freedom of reparametrizations,

\be
h_{\mu \nu}^{\rm polar} = \left(\begin{array}{cccc}
f(r)H_0(r) & H_1(r) & 0 & 0 \\
  &   &   &   \\
H_1(r) & H_2(r)/f(r) & 0 & 0 \\
  &   &   &   \\
0 & 0 & r^2K(r) & 0 \\
  &   &   &   \\
0 & 0 & 0 & r^2K(r) {\rm sin}^2 \theta
\end{array} \right)
e^{-i\omega t}
P_l ({\rm cos} \theta)
\label{pomet}
\ee
by also factorizing the $t$ and $r$ dependence of the four functions $H_0 (t, r)$,
$H_1 (t, r)$, $H_2 (t, r)$ and $K(t, r)$.

In the following, we assume without lose of generality that the metric perturbations
$h_{\mu \nu}$ are decomposed invariantly into axial and polar parts given by the
axisymmetric expressions \eqn{axmet} and \eqn{pomet}. The equations for the
corresponding radial functions $h_0(r)$, $h_1(r)$ and $H_0 (r)$, $H_1 (r)$,
$H_2 (r)$, $K(r)$ follow from the linearized Einstein equations in vacuum,
as will be seen later.

\section{Resolving the duality relations}
\setcounter{equation}{0}

In this section we impose the field equations and reduce Maxwell theory as well as
linear gravity to an effective Schr\"odinger problem, as for all wave equations on
spherically symmetric backgrounds (see also ref. \cite{wald} for a comprehensive but
general discussion). The main result that will be described here is the
manifestation of electric/magnetic duality as axial/polar interchange among field
configurations with opposite parity, following \cite{bakas1}. This resolves the
duality relations and
provides explicit construction of the dual photons and gravitons, in close analogy
to each other, by group theory methods. The results follow by direct computation
without the need to construct explicit solutions of the Schr\"odinger equation; these
solutions will only be presented later in the context of holographic computations in 
$AdS_4$ space-time.

\subsection{Dual photons}

First, it can be shown that Maxwell equations on a spherically symmetric background
with metric
\be
ds^2 = -f(r) dt^2 + {dr^2 \over f(r)} + r^2 (d \theta^2 + {\rm sin}^2 \theta
d \phi^2) ~,
\ee
with arbitrary $f(r)$, are reduced to an effective Schr\"odinger problem, \cite{ruffini},
\cite{wald},
\be
\left(-{d^2 \over dr_{\star}^2} + f(r) {l(l+1) \over r^2} \right)
\Psi (r) = \omega^2 \Psi (r)
\label{mainwave}
\ee
with respect to the tortoise radial coordinate $r_{\star}$, which is defined as
follows,
\be
dr_{\star} = {dr \over f(r)} ~.
\ee

The demonstration will be made separately for the axial and polar sectors
using the decomposition of the gauge field into spherical harmonics of given
parity. The effective wave function will also be identified with appropriate
gauge invariant components of the field $A_{\mu}$, so that electric/magnetic
duality can be readily implemented by exchanging axial with polar sectors.
This way, the explicit construction of $\tilde{A}_{\mu}$ will follow from
$A_{\mu}$ up to gauge transformations.

\underline{Axial sector}: Using the form of the gauge potentials \eqn{axvec},
let us first compute the non-vanishing components of the field strength which 
take the form
\ba
F_{t \phi} & = & -i \omega a_l (r) e^{-i \omega t} {\rm sin} \theta ~
\partial_{\theta} P_l ({\rm cos} \theta) ~, \\
F_{r \phi} & = & a_l^{\prime} (r) e^{-i \omega t} {\rm sin} \theta ~
\partial_{\theta} P_l ({\rm cos} \theta) ~, \\
F_{\theta \phi} & = & - l(l+1) a_l (r) e^{-i \omega t} {\rm sin} \theta ~
P_l ({\rm cos} \theta) ~.
\ea
In terms of physical electric and magnetic fields one sees that the axial
sector has non-vanishing $E_{\phi}$, $B_r$ and $B_{\theta}$.

It follows by straightforward calculation that Maxwell equations
$\nabla_{\nu} F^{\mu \nu} = 0$ reduce to the Schr\"odinger equation \eqn{mainwave}
with the identification of wave-function,
\be
\Psi_{\rm axial} (r) = a_l (r) ~.
\ee

\underline{Polar sector}: In this case, using the form of the gauge potentials
\eqn{povec}, the non-vanishing components of the field strength turn out to be
\ba
F_{tr} & = & -\left(C_l^{\prime} (r) + i \omega D_l (r) \right) e^{-i \omega t}
P_l ({\rm cos} \theta) ~, \\
F_{t \theta} & = & - C_l (r) e^{-i \omega t} \partial_{\theta}
P_l ({\rm cos} \theta) ~, \\
F_{r \theta} & = & - D_l (r) e^{-i \omega t} \partial_{\theta}
P_l ({\rm cos} \theta) ~.
\ea
The non-vanishing components of the physical electric and magnetic fields in the
polar sector are $E_r$, $E_{\theta}$ and $B_{\phi}$, which are complementary
to those of the axial sector; it will shortly be seen that this is not an accident.

As before, Maxwell equations reduce to the Schr\"odinger equation \eqn{mainwave}
provided that the following identification is made for the wave-function
\be
\Psi_{\rm polar} (r) = r^2 \left(C_l^{\prime} (r) + i \omega D_l (r) \right) ~.
\ee
Actually, it also turns out that the non-trivial components of the polar gauge
field \eqn{povec} are expressed in terms of the corresponding wave-function as
\ba
C_l (r) & = & {1 \over l(l+1)} {d \over dr_{\star}} \Psi_{\rm polar} (r) ~, \\
D_l (r) & = & -{i \omega \over l(l+1) f(r)} \Psi_{\rm polar} (r) ~.
\ea

\underline{Duality relation}: It is a simple matter to verify that the two sectors
are mutually related by electric/magnetic duality, i.e.,
\ba
{}^{\star}F_{\mu \nu}^{\rm axial} & = &
F_{\mu \nu}^{\rm polar} ~, \\
{}^{\star}F_{\mu \nu}^{\rm polar} & = &
- F_{\mu \nu}^{\rm axial} ~,
\ea
provided that the same boundary conditions are imposed on the axial and
polar wave-functions so that they are the same (up to an arbitrary factor)
and the frequencies $\omega$ are also the same. With the normalization
chosen above, matching is exact for
\be
\Psi_{\rm polar} = l(l+1) \Psi_{\rm axial} (r) ~.
\ee

Thus, axial/polar interchange provides explicit construction of the dual
photons as advertised above,
\be
\tilde{A}_{\mu}^{\rm axial} = A_{\mu}^{\rm polar} ~, ~~~~~~
\tilde{A}_{\mu}^{\rm polar} = A_{\mu}^{\rm axial} ~.
\ee
This result will be used to guide a similar construction for the dual
gravitons.

\subsection{Dual gravitons}

Next, we examine the linearized Einstein equations on a restricted class of
spherically symmetric backgrounds
\be
ds^2 = -f(r) dt^2 + {dr^2 \over f(r)} + r^2 (d \theta^2 + {\rm sin}^2 \theta
d \phi^2) ~,
\ee
with cosmological constant $\Lambda$ and profile function
\be
f(r) = 1 - {\Lambda \over 3} r^2 ~.
\ee
The field equations $\delta R_{\mu \nu} = \Lambda h_{\mu \nu}$ can also be
reduced to an effective Schr\"odinger problem (see, for instance, \cite{wald} and
\cite{bakas1})
\be
\left(-{d^2 \over dr_{\star}^2} + f(r) {l(l+1) \over r^2} \right)
\Psi (r) = \omega^2 \Psi (r)
\label{mainwave2}
\ee
with respect to the tortoise radial coordinate $r_{\star}$. Note that the effective
potential is the same as in Maxwell theory.

\underline{Axial sector}: Using the form of the metric perturbations \eqn{axmet}
we find that the field equations reduce to the following system of first order
differential equations for the two unknown radial functions $h_0 (r)$ and $h_1 (r)$,
\ba
& & {2 \over r} h_0 (r) - h_0^{\prime} (r) = i {f(r) \over \omega} \left({\omega^2
\over f(r)} - {(l-1)(l+2) \over r^2} \right) h_1 (r) ~, \\
& & h_0 (r) = i {f(r) \over \omega} \left(f(r) h_1 (r) \right)^{\prime} ~.
\ea
These equations give rise to the effective Schr\"odinger equation \eqn{mainwave2}
with wave-function
\be
\Psi_{\rm axial} (r) = {f(r) \over r} h_1 (r) ~.
\ee
Clearly, $h_0 (r)$ is also expressed in terms of $\Psi_{\rm axial} (r)$ as
\be
h_0 (r) = {i \over \omega} {d \over dr_{\star}} (r \Psi_{\rm axial} (r)) ~.
\ee

\underline{Polar sector}: In this case, the metric perturbations assume the form
\eqn{pomet}, which depends on four unknown radial functions $H_0 (r)$, $H_1 (r)$,
$H_2 (r)$ and $K(r)$. The field equations yield the following relation
\be
H_0 (r) = H_2 (r) ~,
\ee
as well as the system of first order equations
\ba
& & r K^{\prime} (r) + {K(r) \over f(r)} - H_0 (r) -
i {l(l+1) \over 2 \omega r} H_1 (r) = 0 ~, \\
& & \left(f(r) H_0 (r) \right)^{\prime} - f(r) K^{\prime} (r)
+ i \omega H_1 (r) = 0 ~, \\
& & \left(f(r) H_1 (r) \right)^{\prime} +
i \omega \left(H_0 (r) + K(r) \right) = 0 ~.
\ea
Furthermore, consistency of the remaining second order field equations requires an
algebraic relation among the remaining three functions,
\ba
& & (l-1)(l+2) H_0 (r) - {2ir \over \omega} \left(\omega^2 + {\Lambda \over 6}
l(l+1) \right) H_1 (r) = \nonumber\\
& & ~~~~~ = \left(l(l+1) - {2 \over f(r)} (\omega^2 r^2 + 1) \right) K(r) ~,
\ea
which can be regarded as first integral of the first order system above.

It can be verified that these coupled differential equations give rise to the
same Schr\"odinger problem \eqn{mainwave2}, as before, provided that the
effective wave-function is constructed as follows,
\be
\Psi_{\rm polar} (r) = {2r \over (l-1)(l+2)} \left(K(r) - i {f(r) \over \omega r}
H_1 (r) \right) .
\ee
Actually, all radial functions can be written in terms of
$\Psi_{\rm polar} (r)$ and its derivative, but the explicit expressions are
more involved now. They read as
\ba
H_0 (r) & = & H_{2} (r) = \left({l(l+1) \over 2r} - {\omega^2 r \over f(r)} +
{d \over dr} \right) \Psi_{\rm polar} (r) ~, \\
H_1(r) & = & -{i \omega \over f(r)} \left(1 + r{d \over dr_{\star}} \right)
\Psi_{\rm polar} (r)  ~, \\
K(r) & = & \left({l(l+1) \over 2r} + {d \over dr_{\star}} \right)
\Psi_{\rm polar} (r) ~. 
\ea

\underline{Duality relation}: It is a very lengthy computation to extract
the electric and magnetic components of the Weyl tensor. The particular
expressions in terms of the effective wave-functions are rather complicated
and will not be included here. However, once
this is done, it is a simple matter to verify that the two sectors
are mutually related by electric/magnetic duality, \cite{bakas1}, i.e.,
\ba
E_{ab}^{\rm polar} & = &
B_{ab}^{\rm axial} ~, \\
B_{ab}^{\rm polar} & = &
- E_{ab}^{\rm axial} ~,
\ea
provided that the same boundary conditions are imposed on the axial and
polar wave-functions so that they are the same (up to an arbitrary factor)
and the frequencies $\omega$ are also the same. Taking into account the
normalization above, the formulae match exactly provided that
\be
\Psi_{\rm axial} = {i \omega \over 2} \Psi_{\rm polar} (r) ~.
\ee

Thus, axial/polar interchange provides explicit construction of the dual
gravitons, in analogy with the construction of dual photons, i.e.,
\be
\tilde{h}_{\mu \nu}^{\rm axial} = h_{\mu \nu}^{\rm polar} ~, ~~~~~~
\tilde{h}_{\mu \nu}^{\rm polar} = h_{\mu \nu}^{\rm axial} ~.
\ee

\section{Holographic implications of duality}
\setcounter{equation}{0}

As application of these results, we describe the boundary manifestation
of electric/magnetic duality for Maxwell theory and linearized gravity on
$AdS_4$ space-time. Since all fields on the bulk are expressed in terms
of the wave-function $\Psi (r)$, their values at the boundary follow by
expanding $\Psi (r)$ as $r \rightarrow \infty$. Thus, we first briefly
discuss the normalizable solution of the Schr\"odinger equation and
its asymptotic expansion under general boundary conditions and then use the
dictionary of AdS/CFT correspondence to obtain the boundary form of duality.
We will end this section with a brief discussion of the energy-momentum/Cotton
tensor duality for $AdS_4$ black-holes.

\subsection{Solving the Schr\"odinger equation}

The effective Schr\"odinger equation that governs the axial and polar sectors of
both Maxwell theory and linearized gravity on $AdS_4$ space-time simplifies to
\be
\left(-{d^2 \over dx^2} + {l(l+1) \over {\rm sin}^2 x} \right) \Psi (x)
= \Omega^2 \Psi (x)
\ee
by first expressing $r$ in terms of $r_{\star}$, as given by
\be
{\rm tan} \left(\sqrt{-{\Lambda \over 3}} ~ r_{\star} \right) =
\sqrt{-{\Lambda \over 3}} ~ r ~,
\ee
and then letting, for convenience,
\be
x = \sqrt{-{\Lambda \over 3}} ~ r_{\star} ~, ~~~~~~
\Omega = \sqrt{-{3 \over \Lambda}} ~ \omega ~.
\ee
Here, $x$ assumes all values from $0$ to $\pi / 2$ as $r$
varies from the origin $r=0$ to spatial infinity $r = \infty$.

This problem can be transformed into a hypergeometric differential equation,
so that the normalizable solution satisfying $\Psi(0) = 0$ at the origin
$r=0$ is provided by
\be
\Psi (x) = {\rm cos} x ~ {\rm sin}^{l+1} x ~ F(a, ~ b; ~ c; ~ {\rm sin}^2 x)
\ee
with coefficients
\be
a = {1 \over 2} \left(l+2 + \Omega \right) , ~~~~
b = {1 \over 2} \left(l+2 - \Omega \right) , ~~~~
c = l+ {3 \over 2} ~.
\ee
Its behavior at spatial infinity is described by the following asymptotic
expansion in powers of $1/r$,
\be
\Psi (r) = I_0 + {I_1 \over r} + {I_2 \over r^2} + {I_3 \over r^3}
+ {I_4 \over r^4} + \cdots ~,
\ee
where the first two coefficient turn out to be
\ba
I_0 & = & \Gamma^{-1} \left({1 \over 2} (l+2 + \Omega)\right)
\Gamma^{-1} \left({1 \over 2} (l+2 - \Omega)\right)  \\
I_1 & = & -2 \sqrt{-{3 \over \Lambda}} ~
\Gamma^{-1} \left({1 \over 2} (l+1 + \Omega)\right)
\Gamma^{-1} \left({1 \over 2} (l+1 - \Omega)\right)
\ea
up to an overall (irrelevant) numerical factor. The remaining coefficients
are determined by $I_0$ and $I_1$.

The boundary conditions at $r = \infty$ are solely expressed
in terms of $I_0$ and $I_1$. Since
\be
I_0 = \Psi (r= \infty) ~, ~~~~~
{\Lambda \over 3} I_1 = {d \Psi \over d r_{\star}} (r = \infty) ~,
\ee
it follows that general boundary conditions (also called mixed or Robin)
can be expressed in terms of the ratio
\be
{I_0 \over I_1} = \gamma
\ee
for fixed constant $\gamma$ that can assume all values, including zero and
infinity. Thus, the allowed spectrum of frequencies $\omega$ obeys a
transcendental relation given by ratios of gamma functions and can only
be solved numerical for general values of $\gamma$. In all cases, however,
the frequencies come in pairs $(\omega, - \omega)$, as can be readily seen
from the particular expressions of $I_0$ and $I_1$ in terms of products of
gamma functions. Consequently, by appropriate superposition of them, real
solutions can always be constructed.

In general, the boundary conditions for the axial and polar sectors
of field theories on $AdS_4$
space-time can be independent from each other, and so is the spectrum of
allowed frequencies, whereas the effective Schr\"odinger equations
are the always the same. Therefore, the axial and polar problems
will be isospectral if and only if both sectors satisfy the same boundary
conditions.

\subsection{Maxwell theory}

We will be brief here, since the holographic manifestation of electric-magnetic 
duality for Maxwell theory is well studied and generally understood as acting on 
the space of three-dimensional (boundary) conformal field theories with $U(1)$ 
symmetry and a chosen coupling to a background gauge field,  \cite{witten} 
(but see also \cite{petkou} and references therein). 

In this case, it is more appropriate 
to define electric and magnetic fields with respect to the radial ADM decomposition,
\be
{\cal E}_a = F_{ra} ~, ~~~~~ {\cal B}_a = {}^{\star}F_{ra} ~, 
\ee
and consider their boundary values as representing the one-point functions of a 
global symmetry current and the curl of the correspondince source, respectively. 
The two are interchanged by axial/polar duality that now acts naturally on the 
moduli space of boundary conformal field theories.

\subsection{Linearized gravity}

We will compute the energy-momentum tensor using holographic renormalization 
(see, for instance, \cite{skenderis1}, \cite{skenderis2} and \cite{kraus}).
For this, we first consider the expression
\be
\kappa^2 T_{ab}^{(r)} = K_{ab} - K \gamma_{ab} -2 \sqrt{-{\Lambda \over 3}}
\gamma_{ab} + \sqrt{-{3 \over \Lambda}} \left(
R_{ab}[\gamma] - {1 \over 2} R[\gamma] \gamma_{ab} \right)
\ee
written in terms of the intrinsic and extrinsic curvature of the
three-geometry $\gamma_{ab}$ that appears at fixed $r$ in the
radial ADM decomposition of the four-dimensional metric. Here,
$\kappa^2 = 8 \pi G$ is the gravitational coupling.
Since the metric acquires an infinite Weyl factor as $r$ is taken to infinity,
it is more appropriate to think of the $AdS_4$ boundary
as a conformal class of boundaries and define the metric on it, $\mathscr{I}$, 
as
\be
ds_{\mathscr{I}}^2 = \lim_{r \rightarrow \infty} \left(-{3 \over \Lambda r^2}
\gamma_{ab} dx^a dx^b \right) .
\ee
Then, the renormalized energy-momentum tensor on $\mathscr{I}$ is defined
accordingly by
\be
T_{ab} = \lim_{r \rightarrow \infty} \left(\sqrt{-{\Lambda \over 3}}
~ r ~ T_{ab}^{(r)} \right)
\ee
and it is finite, traceless and conserved.

When this is applied to linearized gravity in $AdS_4$ space-time, one obtains
explicit expressions for $T_{ab}$ under general boundary conditions. At the same
time, the metric on $\mathscr{I}$, which also depends on the boundary conditions,
is not conformally flat in general. Actually, one may characterize the deviation
from the conformally flat case by computing the Cotton tensor of the boundary
metric, defined as (see, for instance, \cite{cs})
\be
C^{ab} = {1 \over 2 \sqrt{- {\rm det} \gamma}} \left( \epsilon^{acd}
\nabla_c {R^b}_d + \epsilon^{bcd} \nabla_c {R^a}_d \right) ~,
\ee
and obtain explicit expressions for $C_{ab}$ under general boundary conditions.
The details of the calculations will not be presented here, but will rather state
the end result: using the electric and magnetic components of the Weyl tensor, 
which are now taken with respect to the radial ADM decomposition,
\be
{\cal E}_{ab} = Z_{arbr} ~, ~~~~~~ {\cal B}_{ab} = {}^{\star}Z_{arbr} ~,
\ee
the following relation holds true for all boundary conditions,
\ba
& & \lim_{r \rightarrow \infty} \left({\Lambda \over 3} ~ r^3 {\cal E}_{ab} \right)
= \kappa^2 T_{ab} ~, \\
& & \lim_{r \rightarrow \infty} \left({\Lambda^2 \over 9} ~ r^3 {\cal B}_{ab} \right)
= C_{ab}
\ea

Then, we have the following relation among the two distinct type of
perturbations satisfying the same general boundary conditions for $\Psi$, \cite{bakas1},
\ba
C_{ab}^{\rm axial} & = & \kappa^2 T_{ab}^{\rm polar} ~, \\
C_{ab}^{\rm polar} & = & \kappa^2 T_{ab}^{\rm axial} ~,
\ea
which follow from electric/magnetic duality on the bulk and give rise to the
energy-momentum/Cotton tensor duality (also known as dual graviton
correspondence, \cite{other1}, \cite{other2}); see also ref. \cite{other3}
for important earlier work on this subject.

\subsection{Extension to $AdS_4$ black-holes}

We end this section with a brief summary of the situation in the presence
of black holes in $AdS_4$ space-time having mass $m$ and profile metric function
\be
f(r) = 1 - {2m \over r} - {\Lambda \over 3} r^2 ~.
\ee
Here, Newton's constant $G$ is normalized to $1$ for convenience.
In this case there is no electric/magnetic
duality among metric perturbations in the bulk, but there is still an
energy-momentum/Cotton tensor duality at the boundary for appropriately chosen
boundary conditions, as will be seen shortly. Of course, Maxwell equations on
the $AdS_4$ Schwarzschild space-time exhibit electric/magnetic duality that
simply exchanges axial and polar gauge fields, as for all spherically symmetric
backgrounds.

The metric perturbations of $AdS_4$ Schwarzschild solution can still be separated into
axial and polar components with opposite parity, as for $AdS_4$ space-time, based on
spherical symmetry. However, the two sectors do not reduce to the same radial
Schr\"odinger problem, but they rather give rise to two closely related equations,
\cite{wheeler}, \cite{chandra}, \cite{lemos1}, \cite{moss},
\be
\left(- {d^2 \over dr_{\star}^2} + V_{\pm} (r) \right) \Psi_{\pm} (r) =
\omega^2 \Psi_{\pm} (r) ~,
\ee
as in supersymmetric quantum mechanics (see, for instance, \cite{susy}). Here, $(-)$
refers to the axial sector and $(+)$ refers to the polar sector of the theory.
More precisely, it turns out that
\be
V_{\pm} (r) = W^2 (r) \pm {dW (r) \over dr_{\star}} + \omega_{\rm s}^2
\ee
where the superpotential of the partner potentials is
\be
W(r) = {6m f(r) \over r [(l-1)(l+2) r + 6m]} + i \omega_{\rm s}
\ee
with
\be
\omega_{\rm s} = - {i \over 12m} (l-1)l(l+1)(l+2) ~.
\ee
The axial and polar Schr\"odinger problems of black hole perturbations are often
referred in the literature as Regge-Wheeler and Zerilli equations, respectively,
\cite{wheeler}.

Supersymmetric partner potentials have the same energy spectrum (and, hence, the
axial and polar perturbations of the metric have the same frequencies $\omega$)
provided that the corresponding wave-functions satisfy the following first order
relations,
\be
\left(\mp {d \over dr_{\star}} + W(r) \right) \Psi_{\pm} (r) =
i(\omega_{\rm s} \pm \omega)  \Psi_{\mp} (r) ~.
\label{bouconds}
\ee
Then, it is not appropriate to choose the same boundary conditions for
$\Psi_{\pm} (r)$ at $r = \infty$, but rather impose supersymmetric partner
boundary conditions that satisfy equation \eqn{bouconds} at $r = \infty$.
There is a privileged set of such boundary conditions that yield the
static (conformally flat) boundary metric
\be
ds_{\mathscr{I}}^2 = -dt^2 -{3 \over \Lambda} \left(d\theta^2 +
{\rm sin}^2 \theta d\phi^2 \right)
\label{holobound}
\ee
for one type of perturbations and a certain time-dependent (non-conformally flat)
boundary metric for the perturbations with opposite parity. In either case one
finds that the variation of the holographic energy-momentum tensor for
black hole perturbations satisfying the boundary condition \eqn{holobound}
is given by the Cotton tensor of the supersymmetric partner boundary metric,
\cite{bakas}, i.e.,
\be
\kappa^2 \delta T_{ab}^{\rm polar} = C_{ab}^{\rm axial} ~, ~~~~~
\kappa^2 \delta T_{ab}^{\rm axial} = C_{ab}^{\rm polar} ~.
\ee
Thus, although there is no gravitational duality for black hole perturbations, there
is still a dual graviton correspondence at the boundary in certain cases.
The result should be contrasted to the energy-momentum/Cotton tensor
duality for perturbed $AdS_4$ space-time, which, as described above,
is valid for all boundary conditions.

It is also interesting to note in this context that the special boundary conditions
that account for the energy-momentum/Cotton tensor duality for $AdS_4$ black holes
encompass the hydrodynamic modes of very large black holes that saturate the KSS
bound on the ratio of shear viscosity to entropy density, \cite{kss},
\be
{\eta \over s} = {1 \over 4 \pi} ~.
\ee
Further details on these aspects can be found in the published work \cite{bakas}.

\section{Conclusions and discussion}
\setcounter{equation}{0}

We described the electric/magnetic duality of sourceless Maxwell equations
and linearized Einstein equations in parallel. Using the theory of vector and
tensor spherical harmonics, we have provided explicit construction of the dual
photon and graviton configurations by simply exchanging axial and polar field
configurations on spherically symmetric backgrounds. We also considered
the holographic implications of duality for either Maxwell theory or linearized
gravity on $AdS_4$ space-time in the context of $AdS_4/CFT_3$ correspondence.
We expect to have similar duality relations for the two-point correlation
functions of the energy-momentum tensor, as for the one-point functions, but not
for higher correlators, since gravitational duality is only valid at the linear
level. In any case,
the boundary manifestation of gravitational duality as energy-momentum/Cotton
tensor duality should be studied further and also extended to the coupled system
of Einstein-Maxwell field equations. Generalizations of the duality rotations
to higher spin fields were not discussed at all here, but clearly they can
also be investigated using similar techniques based on generalized spherical
harmonics.

There is a closely related result by Fefferman and Graham, \cite{feffe}, which
can be stated as energy-momentum/Cotton tensor {\em self-duality}
in the context of the holographic manifestation of gravitational self-duality.
In particular, self-dual gravitational (instanton) configurations in the bulk
satisfying
\be
{}^{\star} Z_{\mu \nu \rho \sigma} (g) = \pm Z_{\mu \nu \rho \sigma} (g)
\ee
have energy-momentum tensor equal to the Cotton tensor of their boundary metric
$\gamma_{ab}$,
\be
\kappa^2  T_{ab} ({\rm instantons}) = C_{ab} (\gamma) ~.
\ee
$AdS_4$ space-time satisfies trivially this relation, since both sides of the
equation vanish identically, whereas perturbations around it are related by
duality rotations that account for the energy-momentum/Cotton tensor duality
at the boundary. Thus, it is conceivable, that small perturbations of
gravitational instanton configurations will also exhibit duality relations (at
least to linear order) and furthermore there will be a boundary relation of the
form $\kappa^2 \delta T_{ab} = \delta C_{ab}$ among the perturbations of the
holographic energy-momentum tensor and the Cotton tensor of the dual 
perturbed boundary metric. Work in this direction is in progress.

Finally, we end with some ideas about the plausible relevance of ${\cal H}$-space
for our problems. Recall that
the ${\cal H}$-space (also known as {\em heaven}) arose in the literature of
general relativity by considering all shear-free cuts of the complexified
future null infinity of an asymptotically flat Lorentzian four-manifold
satisfying Einstein equations with $\Lambda = 0$ (see, for instance, \cite{ted}
and references therein). In this context, the original surface at infinity
turns into the conformal infinity of a complex four-manifold with a holomorphic
metric whose conformal curvature is self-dual. This construction was
subsequently extended to space-times with a cosmological constant,
\cite{brun}, with the appropriate technical ingredients. It may prove useful
to reformulate the gravitational duality rotations in terms of complex analysis
in ${\cal H}$-space and explore its holographic manifestation (when $\Lambda <0$)
using the self-duality of ${\cal H}$ in the spirit of Fefferman and Graham. 
The entire program of holographic renormalization on $AdS_4$ spaces may very 
well find a useful place in ${\cal H}$-space, turning heaven into powerful calculational 
tool. Work in this direction is also in progress.


\newpage

\begin{thebibliography}{3}
\bibitem{bakas1}
I. Bakas, ``Duality in linearized gravity and holography", Class. Quant.
Grav. \underline{26} (2009) 065013 [arXiv:0812.0152].
\bibitem{bakas}
I. Bakas, ``Energy-momentum/Cotton tensor duality for $AdS_4$ black holes",
JHEP \underline{0901} (2009) 003 [arXiv:0809.4852].
\bibitem{nieto}
J.A. Nieto, ``S-duality for linearized gravity", Phys. Lett. \underline{A262}
(1999) 274 [hep-th/9910049].
\bibitem{hull}
C. Hull, ``Duality in gravity and higher spin gauge fields",
JHEP \underline{0109} (2001) 027 [hep-th/0107149].
\bibitem{claudio}
M. Henneaux and C. Teitelboim, ``Duality in linearized gravity", Phys. Rev.
\underline{D71} (2005) 024018 [gr-qc/0408101].
\bibitem{stan1}
S. Deser and D. Seminara, ``Duality invariance of all free bosonic and
fermionic gauge fields", Phys. Lett. \underline{B607} (2005) 317
[hep-th/0411169].
\bibitem{julia}
B. Julia, J. Levie and S. Ray, ``Gravitational duality near de Sitter space",
JHEP \underline{0511} (2005) 025 [hep-th/0507262].
\bibitem{tassos}
R.G. Leigh and A.C. Petkou, ``Gravitational duality transformations on
$(A)dS_4$", JHEP \underline{0711} (2007) 079 [arXiv:0704.0531].
\bibitem{wheeler}
T. Regge and J.A. Wheeler, ``Stability of a Schwarzschild singularity",
Phys. Rev. \underline{108} (1957) 1063;
F.J. Zerilli, ``Effective potential for even-parity Regge-Wheeler gravitational
perturbation equations", Phys. Rev. Lett. \underline{24} (1970) 737.
\bibitem{zeril}
F.J. Zerilli, ``Gravitational field of a particle falling in a Schwarzschild geometry 
analysed in tensor harmonics", Phys. Rev. \underline{D2} (1970) 2141. 
\bibitem{thorne}
K.S. Thorne, ``Multipole expansions of gravitational radiation",
Rev. Mod. Phys. \underline{52} (1980) 299.
\bibitem{ruffini}
R. Ruffini, J. Tiomno and C.V. Vishveshwara, ``Electromagnetic field of a particle
moving in a spherically symmetric black-hole background", Lett. Nuovo Cim.
\underline{3} (1972) 211.
\bibitem{wald}
A. Ishibashi and R.M. Wald, ``Dynamics on non-globally-hyperbolic static
spacetimes: III. Anti-de Sitter spacetime", Class. Quant. Grav.
\underline{21} (2004) 2981 [hep-th/0402184].
\bibitem{witten}
E. Witten, ``$SL(2, Z)$ action on three-dimensional conformal field theories
with abelian symmetry", [hep-th/0307041].
\bibitem{petkou}
S. de Haro and A.C. Petkou, ``Holographic aspects of electric-magnetic dualities",
J. Phys. Conf. Ser. \underline{110} (2008) 102003 [arXiv:0710.0965].
\bibitem{skenderis1}
S. de Haro, S.N. Solodukhin and K. Skenderis, ``Holographic reconstruction
of space-time and renormalization in AdS/CFT correspondence", Commun. Math.
Phys. \underline{217} (2001) 595
[hep-th/0002230].
\bibitem{skenderis2}
K. Skenderis, ``Asymptotically anti-de Sitter space-times and their stress
energy tensor", Int. J. Mod. Phys. \underline{A16} (2001) 740 [hep-th/0010138];
``Lecture notes on holographic renormalization", Class. Quant. Grav.
\underline{19} (2002) 5849
[hep-th/0209067].
\bibitem{kraus}
V. Balasubramanian and P. Kraus, ``A stress tensor for anti-de Sitter gravity",
Commun. Math. Phys. \underline{208} (1999) 413 [hep-th/9902121].
\bibitem{cs}
S. Deser, R. Jackiw and S. Templeton, ``Topologically massive gauge theories",
Ann. Phys. \underline{140} (1982) 372; Erratum-ibid. \underline{185}
(1988) 406; ``Three-dimensional massive gauge theories", Phys. Rev. Lett.
\underline{48} (1982) 975.
\bibitem{other1}
G. Compere and D. Marolf, ``Setting the boundary free in AdS/CFT", Class. Quant.
Grav. \underline{25} (2008) 195014 [arXiv:0805.1902].
\bibitem{other2}
S. de Haro, ``Dual gravitons in $AdS_4/CFT_3$ and the holographic Cotton tensor",
JHEP \underline{0901} (2009) 042 [arXiv:0808.2054].
\bibitem{other3}
M.T. Anderson, ``$L^2$ curvature and volume renormalizations on AHE metrics on
4-manifolds", Math. Res. Lett. \underline{8} (2001) 171 [math.DG/0011051]. 
\bibitem{chandra}
S. Chandrasekhar, {\em The Mathematical Theory of Black Holes},
Oxford University Press, Oxford, 1983.
\bibitem{lemos1}
V. Cardoso and J.P.S. Lemos, ``Quasinormal modes of Schwarzschild anti-de Sitter
black holes: Electromagnetic and gravitational perturbations", Phys. Rev.
\underline{D64} (2001) 084017 [gr-qc/0105103].
\bibitem{moss}
I.G. Moss and J.P. Norman, ``Gravitational quasinormal modes for anti-de
Sitter black holes", Class. Quant. Grav. \underline{19} (2002) 2323
[gr-qc/0201016].
\bibitem{susy}
F. Cooper, A. Khare and U. Sukhatme, ``Supersymmetry and quantum mechanics",
Phys. Rept. \underline{251} (1995) 267 [hep-th/9405029].
\bibitem{kss}
P.K. Kovtun, D.T. Son and A.O. Starinets, ``Viscosity in strongly interacting
quantum field theory from black hole physics", Phys. Rev. Lett. \underline{94}
(2005) 111601 [hep-th/0405231].
\bibitem{feffe}
C. Fefferman and C.R. Graham, ``The ambient metric" [arXiv:0710.0919].
\bibitem{ted}
M. Ko, M. Ludvigsen and E.T. Newman, ``The theory of ${\cal H}$-space",
Phys. Repts. \underline{71} (1981) 51.
\bibitem{brun}
C.R. LeBrun, ``${\cal H}$-space with a cosmological constant", Proc. R.
Soc. Lond. \underline{A380} (1982) 171.
\end{thebibliography}
\end{document}